\documentstyle[aps,multicol,tighten,epsfig]{revtex}

\begin{document}
\draft 

\title{Synthesis and characterization of entangled mesoscopic superpositions
       \\ for a trapped electron}
\author{Michol Massini,$^{1}$ Mauro Fortunato,$^{1}$ Stefano Mancini,$^{2}$
        and Paolo Tombesi$^{1}$}
\address {$^{1}$INFM and Dipartimento di Matematica e Fisica, Universit\`a di 
Camerino, I--62032  Camerino, Italy \\
$^{2}$INFM and Dipartimento di Fisica, Universit\`a di 
Milano, Via Celoria 16, I--20133  Milano, Italy}
\date{\today}
\maketitle
\begin{abstract}
We propose a scheme for the generation and reconstruction of entangled 
states between the internal and external (motional) degrees of freedom
of a trapped electron. Such states also exhibit quantum coherence at a 
mesoscopic level. 
\end{abstract}


\begin{multicols}{2}

A single electron trapped in a Penning trap~\cite{Brown1} is one of 
the most fundamental quantum systems. Among its peculiar features,
it allows the measurement of fundamental physical constants with striking
accuracy. Recently, for instance, the electron cyclotron degree of freedom
has been cooled to its ground state, where the electron may stay for hours,
and quantum jumps between adjacent Fock states have been observed~\cite{gabr}.
It is therefore evident that the manipulation and the characterization of
the state of a trapped electron is an important issue, with implications in
the very foundations of quantum mechanics.
Earlier works~\cite{rapid,sman} have dealt with this problem for one
(motional) degree of freedom. On the other hand,
{\it entanglement}~\cite{schr} has
been recognized as one of the most puzzling features of quantum mechanics, 
being also the basis of quantum information processing~\cite{quic}.
A striking achievement in this rapidly expanding field has been the recent
entanglement of four trapped ions~\cite{win}. However, it is also 
possible (and conceptually equivalent) to entangle different degrees 
of freedom of the same particle~\cite{win1}.

In the present work we propose to generate {\em entangled} states 
(combined cyclotron and spin states) of an electron in a Penning trap 
by using suitable applied fields. The complete structure of the
cyclotron-spin quantum state is then obtained with the help of a
tomographic reconstruction from the measured data.

In a Penning trap an electron is confined by the combination of a
homogeneous magnetic field along the positive $z$ axis and an
electrostatic quadrupole potential in the $xy$ plane~\cite{Brown1}.
The spatial part of the electronic wave function consists of three degrees
of freedom, but neglecting the slow magnetron motion (whose characteristic
frequency lies in the kHz region), here we only consider the axial
and cyclotron motions, which are two harmonic oscillators radiating in the
MHz and GHz regions, respectively. On the other hand, the spin dynamics
results from the interaction between the magnetic moment of the electron
and the static magnetic field, so that the free Hamiltonian reads
as~\cite{Brown1}
\begin{equation}
\label{Hfree}
{\hat H}_{\rm free}=\hbar\omega_z {\hat a}_z^{\dag} {\hat a}_z
+\hbar\omega_c {\hat a}_c^{\dag} {\hat a}_c
+\hbar\omega_s{\hat \sigma}_z/2\,,
\end{equation}
where the indices $z$, $c$, and $s$ refer to the axial, cyclotron and spin
motions, respectively.

Here, in addition to the usual trapping fields, we consider an external
radiation field as a standing wave along the $z$ direction and rotating,
{\it i.e.} circularly polarized, in the $xy$ plane with frequency
$\Omega$~\cite{mmt}. To be more specific, we consider a standing wave within
the cylindrical cavity configuration~\cite{tan} with the (dimensionless)
wave vector $\kappa$. Then, the interaction Hamiltonian~reads \cite{mmt}
\begin{eqnarray}
\label{Hnodip}
{\hat H}_{\rm int}&=&
\hbar\epsilon\left[{\hat a}_c e^{i\Omega t}
+{\hat a}_c^{\dag}e^{-i\Omega t}\right]
\cos(\kappa\hat{z} +\phi)
\nonumber\\
&+&\hbar\zeta\left[
{\hat\sigma}_-e^{i\Omega t}
+{\hat\sigma}_+e^{-i\Omega t}
\right]
\sin(\kappa \hat{z} +\phi)\;,
\end{eqnarray}
where
${\hat\sigma}_{\pm}=({\hat\sigma}_x \pm i{\hat\sigma}_y)/2$, and
$\hat{z}=\hat{a}_{z}+\hat{a}_{z}^{\dagger}$.
The phase $\phi$ defines the position of the center of the axial motion
with respect to the wave. Depending on its value the electron can be
positioned in any place between a node ($\phi=0$) and an antinode
($\phi=\pm\pi/2)$. The two coupling constants $\epsilon$ and $\zeta$ 
are proportional to the amplitude of the applied radiation field.
For our purposes we also consider the possibility to introduce pulsed
standing waves~\cite{mil} through the microwave inlet~\cite{Brown1}
so that $\epsilon$ and $\zeta$ become time dependent.
The duration of the pulse is assumed to be much shorter than the
characteristic axial period, which is of the order of microseconds.
Depending on $\Omega$ and $\phi$, the interaction Hamiltonian~(\ref{Hnodip})
gives rise to different contributions at leading order in the Taylor 
expansion of $\sin(\kappa \hat{z} +\phi)$ and $\cos(\kappa\hat{z} 
+\phi)$.

Now, nonclassical cyclotron states can be entangled with the spin states
through the following steps. First, we consider $\phi=0$, $\Omega=\omega_s$,
and a pulsed standing wave of duration $\Delta t_1=t_{1}-t_{0}=t_{1}$.
In the following we shall work in a frame rotating at the
frequency $\omega_{s}$. Then, the total Hamiltonian ($\hat{H}_{\rm 
free} + \hat{H}_{\rm int}$) can be written as
$\hat{H}=\hbar\omega_z {\hat a}_z^{\dag} {\hat a}_z
+\hbar(\omega_c -\omega_{s}){\hat a}_c^{\dag} {\hat a}_c
+\hbar\zeta(t){\hat\sigma}_x \kappa ({\hat a}_{z}+\hat{a}_{z}^{\dagger})$.
The first two terms can be neglected during the pulse duration.
In fact, the latter is assumed to be much smaller than $\omega_{z}^{-1}$ and 
$(\omega_{c}-\omega_{s})^{-1}\approx \omega_{z}^{-1}$~\cite{Brown1}. 
Its effect on the axial degree of freedom can be described by means of the
relation ${\hat p}_z(t_1)={\hat p}_z(0)-{\tilde\zeta}{\hat\sigma}_x$,
where ${\tilde\zeta}=\kappa \, \int_0^{t_1} \,dt\,\zeta(t)$, and
$\hat{p}_{z}=-i(\hat{a}_{z}-\hat{a}_{z}^{\dagger})$.
Subsequently, we allow a free evolution for a time $\Delta 
t_2=t_{2}-t_{1}=\pi/(2\omega_z)$.
That amounts to having ${\hat z}(t_2)={\hat p}_z(t_1)={\hat 
p}_z(0)-{\tilde\zeta}{\hat\sigma}_x$.

Finally, we consider the action of another pulsed standing wave
with $\phi=-\pi/2$, $\Omega=\omega_c$, for a time $\Delta t_3=t_{3}-t_{2}$. 
In such a case the  effective Hamiltonian (in the frame rotating at 
the frequency $\omega_{s}$) becomes
${\hat H}=\hbar\omega_z {\hat a}_z^{\dag} {\hat a}_z
+ \hbar(\omega_{c}-\omega_{s})\hat{a}_{c}^{\dagger} \hat{a}_{c}
+ \hbar\epsilon(t)\left({\hat a}_c
+ {\hat a}_c^{\dag}\right) \kappa {\hat z}$. Again, since 
$\omega_{c}-\omega_{s}$ is of the order of $\omega_{z}$~\cite{Brown1},
we can neglect the first two terms in the previous Hamiltonian.
Then, the time evolution operator is equivalent 
to a displacement operator~\cite{gla}
${\hat U}(\Delta t_3)={\hat D}\left(-i{\tilde\epsilon} {\hat z}\right)$,
with
${\tilde\epsilon}=\kappa \, \int_{t_{2}}^{t_3} \,dt\,\epsilon(t)$.
Since ${\hat z}$ is not affected by the time evolution under the previous
Hamiltonian, it remains unaltered during $\Delta t_{3}$, that is
${\hat U}(t_3-t_{0})=\exp\left\{
-i{\tilde\epsilon} ({\hat a}_c+{\hat a}^{\dag}_c)
\left[{\hat p}_z-{\tilde\zeta} {\hat\sigma}_x \right]\right\}$.
This means that the state evolution for the spin-cyclotron system is
given by
\begin{eqnarray}
{\hat\rho}(t_{3}) & = & {\rm Tr}_z\left\{
{\hat D}\left(-i{\tilde\epsilon}{\hat p}_z\right)
{\hat D}\left(i{\tilde\epsilon}{\tilde\zeta}{\hat\sigma}_x\right)
\right.
\nonumber \\
& & \times \left.{\hat R}(0)
{\hat D}^{\dag}\left(i{\tilde\epsilon}{\tilde\zeta}{\hat\sigma}_x\right)
{\hat D}^{\dag}\left(-i{\tilde\epsilon}{\hat p}_z\right)
\right\}\;,
\label{rhot3}
\end{eqnarray}
where ${\hat R}(0)$ is the initial density operator for the whole system.
If we consider for instance the initial axial state as 
a Gaussian state with momentum width $d$, the above equation 
can be rewritten as
\begin{eqnarray}
{\hat\rho}(t_{3}) & = & \frac{1}{\pi^{1/4} d}\int\,dp_z\,
{\hat D}\left(-i{\tilde\epsilon}p_z\right)
{\hat D}\left(i{\tilde\epsilon}{\tilde\zeta}{\hat\sigma}_x\right)
\nonumber \\
 & & \times {\hat\rho}(0)
{\hat D}^{\dag}\left(i{\tilde\epsilon}{\tilde\zeta}{\hat\sigma}_x\right)
{\hat D}^{\dag}\left(-i{\tilde\epsilon}p_z\right)
\,\exp(-p_z^2/d^2)
\;.
\label{rhofin}
\end{eqnarray}
Assuming ${\tilde\zeta}\gg d$, which is easily obtained in the case of
the ground state of the axial oscillator, we can approximate
the evolution of an initial pure state
${\hat\rho}(0)=|\Phi_0\rangle\langle\Phi_0|$
into $|\Phi\rangle= {\hat D}(\alpha{\hat\sigma}_x) |\Phi_0\rangle$,
where $\alpha=i{\tilde\epsilon} {\tilde\zeta}$.

It is now immediate to see that an initial state $|\Phi_0\rangle=
|0\rangle|\uparrow\rangle$
evolves into
\begin{equation}\label{final}
|\Phi\rangle=\left( |\uparrow\rangle |\alpha_+\rangle
+|\downarrow\rangle |\alpha_-\rangle \right)/\sqrt{2}\,,
\end{equation}
where $|\uparrow\rangle$ and $|\downarrow\rangle$ denote spin 
eigenstates, while
$|\alpha_{\pm}\rangle={\cal N}(|\alpha\rangle \pm |-\alpha\rangle)$ 
are the even-odd coherent states~\cite{dod} of the cyclotron mode
with ${\cal N}$ a normalization factor.
Since the latter are orthogonal states, Eq.~(\ref{final}) represents a
maximally entangled state~\cite{maxent}.
On the other hand, even and odd coherent states may represent the 
basis of unconventional quantum bits~\cite{coc}, hence
states of the form of Eq.~(\ref{final}) could be of importance
to encoding and manipulating quantum information~\cite{quic}. Furthermore,
a simple spin rotation (as it can be seen below) is sufficient to realize
the transformation
\begin{equation}
|\Phi \rangle \rightarrow \left( |\alpha\rangle |\uparrow\rangle 
+|-\alpha\rangle |\downarrow\rangle \right)/\sqrt{2}\,,
\label{eq:cat}
\end{equation}
a state already discussed in Refs.~\cite{win1,haroche}. The electronic
state~(\ref{eq:cat}) possesses two very interesting features: first,
if $|\alpha|$, i.e. the product ${\tilde\epsilon}{\tilde\zeta}$ is 
much larger than 1, $|\Psi\rangle$ is a typical example of
{\it Schr\"odinger-cat state}~\cite{schr,zur}.
Second, the full state of the trapped electron is an {\it entangled state}
between the spin and cyclotron degrees of freedom.
It is worth noting that states like~(\ref{final}) or (\ref{eq:cat})
persist for many cycles since the decoherence ({\it i.e.} the rapid 
destruction of superposition states due to the entanglement with the 
environment~\cite{zur}) in such a system is quite
small~\cite{Brown1} (in contrast, in Ref.~\cite{rapid}
the mesoscopic superposition appears only cyclically).

The most general pure state of the trapped electron can be cast in the 
form $|\Psi\rangle = c_{1}|\psi_{1}\rangle|\uparrow\rangle + 
c_{2}|\psi_{2}\rangle|\downarrow\rangle \;,$
$|\psi_{1}\rangle$ and $|\psi_{2}\rangle$ being two unknown cyclotron
states, and the complex coefficients $c_{1}$ and $c_{2}$ satisfying the 
normalization condition $|c_{1}|^{2}+|c_{2}|^{2}=1$.
The density operator $\hat{\rho}=|\Psi\rangle\langle\Psi |$
associated to the pure state $|\Psi\rangle$ can be expressed in the form
\begin{equation}
{\hat\rho} = \left( \begin{array}{cc} 
|c_{1}|^{2}|\psi_{1}\rangle\langle\psi_{1}| &
c_{1}c_{2}^{*}|\psi_{1}\rangle\langle\psi_{2}| \\  &  \\
c_{2}c_{1}^{*}|\psi_{2}\rangle\langle\psi_{1}| &
|c_{2}|^{2}|\psi_{2}\rangle\langle\psi_{2}| \end{array} \right) \;,
\label{eq:Rmat}
\end{equation}
whose elements
$\rho^{(ij)}=c_{i}c_{j}^{*}|\psi_{i}\rangle\langle \psi_{j}|$
are operators in the cyclotron Hilbert space.
The phase-space description corresponding to~(\ref{eq:Rmat}) is given 
by the Wigner-function matrix~\cite{wig,wal} whose elements are
\begin{equation}
\tilde{W}_{ij}(\alpha) = \langle \hat{\delta}_{ij} (\alpha -\hat{a})\rangle
={\rm Tr}[\hat{\rho}\,\hat{\delta}_{ij}(\alpha -\hat{a})]\;,
\label{eq:wij}
\end{equation}
where $\hat{\delta}_{ij}(\alpha -\hat{a})$
is the Fourier transform of the displacement operator
with $i,j=1,2$ denoting the spin components.

In order to characterize the generic state~(\ref{eq:wij}) 
we use a simple reconstruction procedure: Adding 
a particular inhomogeneous magnetic field---known as the ``magnetic bottle''
field~\cite{Brown1}---to that already present in the trap, it is 
possible to perform a simultaneous measurement of both the spin and 
the cyclotronic excitation numbers. The useful interaction Hamiltonian 
for the measurement process is then~\cite{Brown1} 
\begin{equation}
\hat{H}_{\rm bottle} = \hbar \omega_{b} \left[\hat{a}^{\dagger}_{c}
\hat{a}_{c} + 
\frac{g}{2}\frac{\hat{\sigma}_{z}}{2} \right] \hat{z}^{2}\;,
\label{eq:hint}
\end{equation}
where the angular frequency $\omega_{b}$ is directly related to the 
strength of the magnetic bottle field.

Eq.~(\ref{eq:hint}) describes the fact that the axial angular 
frequency is affected both by the number of cyclotron excitations 
$\hat{n}_{c}=\hat{a}^{\dagger}_{c}\hat{a}_{c}$ and by the eigenvalue 
of $\hat{\sigma}_{z}$. 
The modified (shifted) axial frequency can be experimentally
measured~\cite{Brown1} after the application of the inhomogeneous magnetic
bottle field. One immediately sees that it assumes different values for every
pair of eigenvalues of $\hat{n}_{c}$ and $\hat{\sigma}_{z}$, due to the fact
that the electron $g$ factor is slightly (but measurably~\cite{Brown1})
different from 2. Then, repeated measurements of this type allow us to recover
the probability amplitudes associated to the two possible spin states and the
cyclotron probability distribution in the Fock basis. The reconstruction of 
the density matrices $|\psi_{i}\rangle\langle\psi_{i}|$ ($i=1,2$)
in the Fock basis is then possible by employing a technique similar to 
the Photon Number Tomography (PNT)~\cite{sman,man} which exploits a 
phase-sensitive reference field that displaces in the phase space the
particular state one wants to reconstruct~\cite{mafo}.

Following Ref.~\cite{sman}, immediately before the measurement, we apply
a pulsed standing wave~(\ref{Hnodip}) tuned to $\Omega=\omega_c$
with $\phi=0$ in order to get a displacement 
$\gamma=-i\tilde{\epsilon}/\kappa$ on the cyclotron.
Thus we can interpret the quantity
\begin{eqnarray}
P^{(i)}(n,\gamma) & = &
 \langle n| \hat{D}^{\dagger}(\gamma) \hat{\rho}^{(ii)}
 \hat{D}(\gamma) |n\rangle
\nonumber \\
 & = & \langle n,\gamma | \hat{\rho}^{(ii)} |n,
 \gamma\rangle\;,
\label{eq:pnc}
\end{eqnarray}
as the probability of finding the cyclotron state $|\psi_{i}\rangle$ 
in a displaced number state $|n,\gamma\rangle$~\cite{knight}. It should
be noted at this stage that the probability distribution~(\ref{eq:pnc})
is not normalized to unity. Instead, one has
\begin{equation}
	\sum_{n=0}^{\infty} P^{(i)}(n,\gamma)=|c_{i}|^{2}\;.
	\label{ci}
\end{equation}
Eq.~(\ref{ci}) allows us to retrieve in a simple way the moduli of the
coefficients $c_{1}$ and $c_{2}$ in Eq.~(\ref{eq:Rmat}) from the measured
data. Fixing a particular value of $\gamma$, it is then possible to recover
the probability distribution~(\ref{eq:pnc}) performing many identical
experiments.

Expanding the density operator $\hat{\rho}^{(ii)}$ in the Fock basis, 
and defining $N_{c}$ as an appropriate estimate of the maximum number 
of cyclotronic excitations (cut-off), we have
\begin{equation}
P^{(i)}(n,\gamma) = \sum_{k,m=0}^{N_{c}} \langle n, \gamma
|k\rangle\langle k|\hat{\rho}^{(ii)}|m\rangle\langle 
m|n,\gamma\rangle\;.
\label{eq:pinc}
\end{equation}
The projection of the displaced number state $|n,\gamma\rangle$ 
onto the Fock state $|m\rangle$ can be obtained generalizing the 
result derived in Ref.~\cite{cah}.

Let us now consider, for a given value of $|\gamma|$, $P^{(i)}(n,\gamma)$
as a function of $\varphi=\arg[\gamma]$ \cite{opa} 
and calculate the coefficients of the Fourier expansion
\begin{equation}
P^{(i)} (n,s) = \frac{1}{2\pi}\int_{0}^{2\pi} d\varphi \;
P^{(i)}(n,\varphi) e^{is\varphi}\;,
\label{eq:psina}
\end{equation}
for $s=0,1,2,\ldots$. Combining Eqs.~(\ref{eq:pinc}) and 
(\ref{eq:psina}), we get
\begin{equation}
P^{(i)} (n,s) = \sum_{m=0}^{N_{c}-s} {G}^{(s)}_{n,m}
(|\gamma|) \langle m+s | \hat{\rho}^{(ii)} | m \rangle\;,
\label{eq:psinag}
\end{equation}
where the explicit expression of the matrices $G$ is given in
Ref.~\cite{michol}.

We may now note that if the distribution $P^{(i)}(n,\gamma)$ is 
measured for $n\in [0,N]$ with $N\geq N_{c}$, then 
Eq.~(\ref{eq:psinag}) represents for each value of $s$ a system of 
$N+1$ linear equations between the $N+1$ measured quantities and the 
$N_{c}+1-s$ unknown density matrix elements. Therefore, in order to 
obtain the latter, we only need to invert the system
\begin{equation}
\langle m+s | \hat{\rho}^{(ii)} | m \rangle = \sum_{n=0}^{N}
M^{(s)}_{m,n} (|\gamma|) P^{(i)} (n,s)\;,
\label{eq:msrhom}
\end{equation}
where the matrices $M$ are given by $M=(G^{T} G)^{-1} G^{T}$.
Since the overdetermined system~(\ref{eq:psinag}) is inverted using 
the method of least squares, we are sure that when the measured 
probabilities are slightly inaccurate, the quantities calculated from 
the reconstructed density matrix best fit the measured 
ones~\cite{opa}.

However, this kind of measurement does not allow to retrieve the 
relative phase $\theta$ between the complex coefficients $c_{1}$ and 
$c_{2}$. We can then again use the Hamiltonian~(\ref{Hnodip}) 
tuned to have a spin rotation, {\it i.e.} $\Omega=\omega_s$ and $\phi=\pi/2$.
After a $\pi/2$ spin rotation we have
\begin{eqnarray}
|\Psi\rangle\to|{\overline\Psi}\rangle
 & = & \frac{\protect\sqrt{2}}{2}
 \left[(c_{1}|\psi_{1}\rangle - ic_{2}|\psi_{2}\rangle)|\uparrow\rangle\right.
\nonumber \\
 & & \left. + (-ic_{1}|\psi_{1}\rangle
 +c_{2}|\psi_{2}\rangle)|\downarrow\rangle\right]\;.
\label{eq:psibar}
\end{eqnarray}
We can now repeat the spin measurements just as we have described 
above in the case of the {\em unknown} initial state $|\Psi\rangle$.
Repeating this procedure over and over again (with the same unknown 
initial state) for a large number of times and tracing out the cyclotron
degree of freedom (no drive in this case is required), it is possible to
recover the probabilities $\bar{P}^{(i)}$ associated to the two spin
eigenvalues for the state $|{\overline\Psi}\rangle$. Without loss of 
generality, we can assume $c_{1}\in {\bf R}$, $c_{2}=|c_{2}| e^{i\theta}$, 
and $\langle\psi_{1}|\psi_{2}\rangle = re^{i\beta}$, which yield
\begin{mathletters}
\label{eq:pbar}
\begin{eqnarray}
\bar{P}^{(1)} & = & \frac{1}{2}[1+2r|c_{1}| |c_{2}| 
\sin(\theta+\beta)]
\label{eq:pbarup} \\
\bar{P}^{(2)} & = & \frac{1}{2}[1-2r|c_{1}| |c_{2}| 
\sin(\theta+\beta)]\;.
\label{eq:pbardown}
\end{eqnarray}
\end{mathletters}
It is important to note that the probabilities $\bar{P}^{(i)}$  
can be experimentally sampled and that the  modulus $r$ and the phase
$\beta$ of the scalar product $\langle\psi_{1}|\psi_{2}\rangle$ can be
both derived from the reconstruction of the cyclotron density matrices
$\rho^{(11)}$ and $\rho^{(22)}$~\cite{michol}.
Thus we are able to find the relative phase $\theta$ by simply
inverting one of the two Eqs.~(\ref{eq:pbar}), {\it e.g.}
\begin{equation}
\theta = \arcsin \left[\frac{2\bar{P}^{(1)}-1}{2r |c_{1}| |c_{2}|}
\right]- \beta\;,
\label{eq:theta}
\end{equation}
where the ambiguity in the $\arcsin$ function can be removed by 
repeating the procedure above using a second spin rotation~(\ref{eq:psibar})
with an angle different from $\pi/2$. As in any tomographic scheme, our 
reconstruction procedure yields the state to be measured up to an
uninteresting phase factor which, however, does not affect the results.

\begin{figure}[t]
\centerline{\epsfig{figure=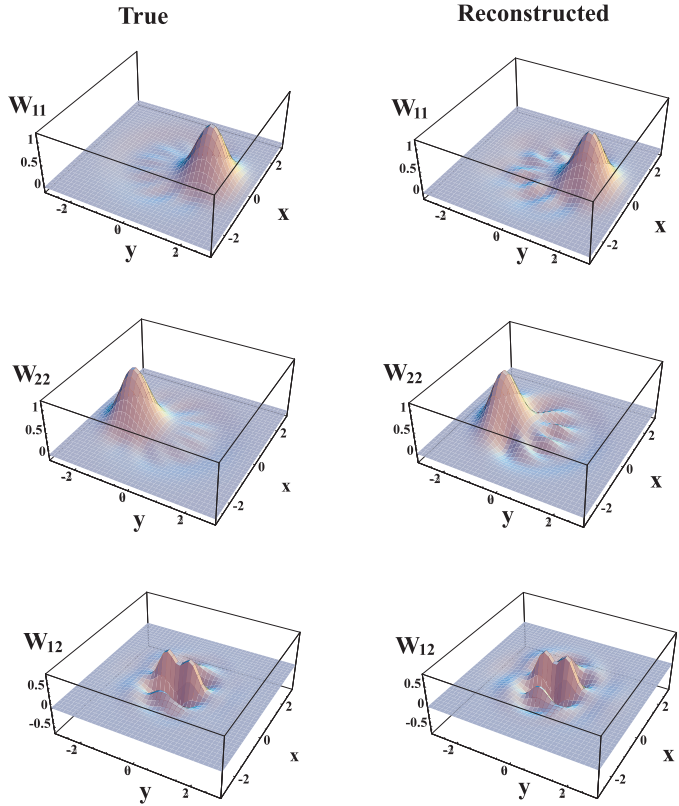,width=3.5in}}
\vspace{0.1cm}
\caption{\narrowtext Simulated tomographic reconstruction of the
Wigner matrix $W_{ij}=\tilde{W}_{ij}/c_{i}c_{j}^{*}$ for the state of
Eq.~(\ref{eq:cat}) with $\alpha=3i/2$. The quantum efficiency is
$\eta=0.9$ and 10$^{6}$ data per phase have been simulated. In this
simulation an amplitude $|\gamma|=1.2$ of the applied reference field
has been used. The theoretical distributions are plotted on the left 
for comparison.}
\label{fg:one}
\end{figure}

As an example of the states that can be generated with the method 
outlined above, and of the application of the proposed reconstruction
procedure, we show in Fig.~\ref{fg:one} the results of numerical Monte-Carlo 
simulations of the reconstruction of the
state~(\ref{eq:cat}). In this simulation we have used the value $|\alpha|=1.5$
which  is experimentally accessible~\cite{Brown1} and gives {\it mesoscopic}
entangled superpositions of cyclotron coherent states with opposite 
phase. In order to account for actual experimental conditions, we 
have also considered the effects of a non-unit quantum efficiency $\eta$ in 
the counting of cyclotron excitations. When $\eta < 1$, the actually 
measured distribution is related to the ideal distribution by a binomial
convolution~\cite{scula}. As it can be seen from Fig.~\ref{fg:one}, the
reconstructed distributions turn out to be quite faithful. We also would
like to emphasize that the particular shape of $W_{12}$ is due to the
quantum interference given by the entanglement between the two degrees
of freedom: in fact, in absence of entanglement $\rho^{(12)}$ would just
be a replica of the diagonal parts $\rho^{(11)}$ and $\rho^{(22)}$.

We have performed a large number of simulations with different states 
and several values of the parameters, which confirm that the present 
method is quite stable and accurate. In addition, and for all the 
cases considered, the values of the parameters $c_{1}$, $c_{2}$, and 
$\theta$ are very well recovered, with a relative error of the order
of $10^{-5}$.

To conclude, we have proposed here a method which is able to synthesize
and characterize highly nonclassical, maximally entangled states for a single
trapped electron.
The method is based on simple operations (switching on and off standing waves)
which are currently realized in the present trap technology~\cite{gabr}.
An experimental implementation of the proposed method might yield new insight
in the foundations of quantum mechanics and allow further progress in the 
field of quantum information.

We gratefully thank G.~M.~D'Ariano for help with the numerical 
simulations. This work has been partially supported by INFM, by the
TMR Network ``Microlasers and Cavity QED'', and by MURST.

%\end{multicols}

%\begin{multicols}{2}

\end{multicols}

%\widetext

\end{document}